\def\year{2017}\relax
\documentclass[letterpaper]{article} 
\usepackage{aaai17}  
\usepackage{times}  
\usepackage{helvet}  
\usepackage{courier}  
\usepackage{url}  
\usepackage{graphicx}  
\usepackage{algorithm}
\usepackage[noend]{algpseudocode}
\begin{document}
%
\title{Reinforcement Learning based Embodied Agents Modelling Human Users Through Interaction and Multi-Sensory Perception}
\author{Kory W. Mathewson, Patrick M. Pilarski\\
Departments of Computing Science and Medicine\\
University of Alberta, Edmonton, Alberta, Canada\\
{[}korym, pilarski{]} @ ualberta.ca\\
}
\maketitle
\begin{abstract}
This paper extends recent work in interactive machine learning (IML) focused on effectively incorporating human feedback. We show how control and feedback signals complement each other in systems which model human reward. We demonstrate that simultaneously incorporating human control and feedback signals can improve interactive robotic systems’ performance on a self-mirrored movement control task where a RL-agent controlled right arm attempts to match the pre-programmed movement pattern of the left arm. We illustrate the impact of varying human feedback parameters on task performance by investigating the probability of giving feedback on each time step and the likelihood of given feedback being correct. We further illustrate that varying the temporal decay with which the agent incorporates human feedback has a significant impact on task performance. We found that \textit{smearing} human feedback over time steps improves performance and we show varying the probability of feedback at each time step, and an increased likelihood of those feedbacks being 'correct', can impact agent performance. We conclude that understanding latent variables in human feedback is crucial for learning algorithms acting in human-machine interaction domains.
 
\end{abstract}

\section{Introduction}
Reinforcement learning (RL) agents can learn optimal actions through building models of environments through perceptive sensors during repeated interactions. Often RL agents cooperate interactively with human trainers to solve difficult tasks. Human teachers are a unique component of the environment who may deliver control signals and contextual information through feedback. As human-robot interaction becomes more complex, due to rapid advancements in actuator and sensor technology, a significant gap emerges between the number of possible control signals a human can provide and the number of controllable actuators a robotic system. There is often a limited set of control signals which a human can provide, and a large number of robotic system controllable functions. 

The limit of human provided control signals is of particular interest in the field of robotic prostheses—artificial limbs attached to the body to augment and/or replace abilities lost through injury or illness. 
\begin{figure}[ht]
	\centering
    \includegraphics[width=3.3in]{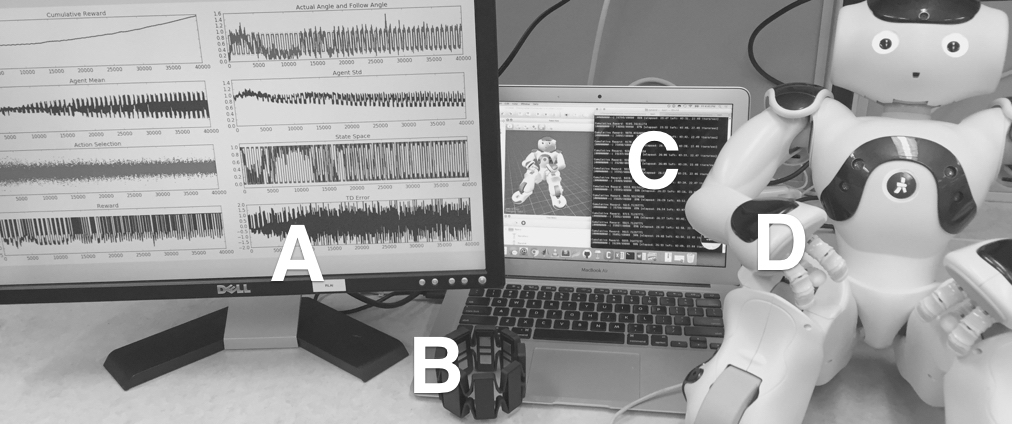}
  	\caption{Configuration with A) results example, B) \textit{Myo}, C) simulation/learning/feedback System, and D) \textit{Nao}.}
    \label{fig:fig1}
\end{figure}
Prosthetic limbs with many degrees-of-freedom have been developed \cite{Castellini2014ProceedingsElectromyography}. State-of-the-art prosthetics can perform complex functions and movements, but rapid, reactive control of this functionality, by human users, is limited; this limitation causes some users to abandon their devices \cite{Castellini2014ProceedingsElectromyography,Biddiss2007UpperYears.,Micera2010ControlInformation,Scheme2011ElectromyogramUse}. New methods are in development to help humans control complex robotic devices through intelligent control sharing and by allowing a learning agent inside the prosthetic to model the human user. The work presented herein explores RL agents controlled by simulated human electromyography (EMG) signals, with additional reward feedback signals.

\section{Background}
RL is a learning framework inspired by behaviorism \cite{Skinner1938TheAnalysis} which describes how agents improve over time by taking actions in an environment with a goal of maximizing \textit{expected return}, the cumulative future reward signal received by the agent \cite{Sutton1998ReinforcementIntroduction}. An agent’s control policy is iteratively improved by selecting actions which maximize \textit{return}. RL problems are often described as sequential decision making problems modelled as Markov Decision Processes (MDPs) which define tuples: $(State, Action, Transitions, \gamma, Reward)$, full details of MDPs are omitted for space and can be found in \cite{Sutton1998ReinforcementIntroduction,Mathewson2016SimultaneousLearning}. The ultimate goal of an RL agent is to determine a policy which maps a given current state to the correct actions to maximize \textit{expected return}. In this work we use a continuous actor-critic (AC) algorithm (Algorithm \ref{alg:cacrl}) similar to that described in \cite{Pilarski2011OnlineLearning,Pilarski2013Real-timeJoints,Mathewson2016SimultaneousLearning}. AC methods can reduce variance in gradient estimation through the use of two learning systems: a policy-focused actor (selects the best action) and a critic (estimate of value function, criticizes actor) \cite{Sutton1998ReinforcementIntroduction}.

The \textit{Interactive Shaping Problem (ISP)} defines the problem of optimizing the incorporation of human feedback into a learning agent in a sequential decision making problem \cite{Knox2010CombiningLearning}. The \textit{ISP} asks: how can the agent learn the best possible task policy as measured by task performance or cumulative human feedback, given the information contained in the human responses \cite{Knox2009InteractivelyFramework,Knox2012ReinforcementTasks}. While there are many ways to incorporate human knowledge into a learning system both before and during learning \cite{Thomaz2008TeachableLearners,Chernova2014RobotTeachers}, this paper focuses on incorporating human feedback directly alongside MDP derived reward.

This work builds on the work of Vien and Ertel, who showed that the human feedback model can be generalized to address the problems associated with periods of noisy, and/or inconsistent, human feedback \cite{Vien2013LearningSpaces}. Recent advancements in modelling human feedback with a Bayesian approach have improved on the work of Knox and Stone in discrete environments \cite{Loftin2016LearningLearning}. Most recently work by \citeauthor{Macglashan2016ConvergentHumans} show that human feedback may be better modelled as an advantage function to handle changes in a human’s feedback strategy over time \cite{Macglashan2016ConvergentHumans}.

In this study, we explore the implications of varying several latent variables in human feedback for learning algorithms acting in complex human-machine interaction domains. We investigate the probability of the human trainer providing feedback, the probability that feedback is correct, and the effect of \textit{smearing} that feedback over time to account for the limited number of time steps with direct human feedback.

\begin{algorithm}
\caption{Continuous Actor-Critic Reinforcement Learning}\label{alg:cacrl}
\begin{algorithmic}[1]
\State \textbf{initialize:} $\mathbf{w}_\mu, \mathbf{w}_\sigma, \mathbf{v}, \mathbf{e}_\mu, \mathbf{e}_\sigma, \mathbf{e}_\mathbf{v}, s$
\Repeat
\State $\mu \leftarrow \mathbf{w}_\mu^T \mathbf{x}(s)$
\State $\sigma \leftarrow \exp[\mathbf{w}_\sigma^T \mathbf{x}(s)]$
\State $a \leftarrow \mathcal{N}(\mu,\sigma^2)$
\State \textbf{take action} $a$, \textbf{observe} $r, s'$
\State $\delta \leftarrow r + \gamma\mathbf{v}^T \mathbf{x}(s') - \mathbf{v}^T \mathbf{x}(s)$
\State $\mathbf{e}_\mathbf{v} \leftarrow \min[1,\lambda_\mathbf{v} \gamma \mathbf{e}_\mathbf{v} + \mathbf{x}(s)]$
\State $\mathbf{v} \leftarrow \mathbf{v} + \alpha_\mathbf{v} \delta \mathbf{e}_\mathbf{v}$
\State $\mathbf{e}_\mathbf{\mu} \leftarrow \lambda_\mathbf{w} \mathbf{e}_\mu + (a - \mu) \mathbf{x}(s)$
\State $\mathbf{w}_\mu \leftarrow \mathbf{w}_\mu + \alpha_\mu \delta \mathbf{e}_\mu$
\State $\mathbf{e}_\mathbf{\sigma} \leftarrow \lambda_\mathbf{w} \mathbf{e}_\sigma + [(a - \mu)^2 - \sigma^2] \mathbf{x}(s)$
\State $\mathbf{w}_\sigma \leftarrow \mathbf{w}_\sigma + \alpha_\sigma \delta \mathbf{e}_\sigma$
\State $s \leftarrow s'$
\Until{termination criteria is met}
\end{algorithmic}
\end{algorithm}

\section{Methods}
\subsection{Aldebaran Nao and Myo EMG Data}
The experimental set up is shown in \ref{fig:fig1}. It is composed of the Aldebaran Nao robotic platform (Aldebaran Robotics), a wireless Myo EMG armband (Thalmic Labs), and a MacBook Air (Apple, 2.2 GHz Intel Core i7, 8GB RAM) for human feedback and running the learning agent.

The experiments in this paper are performed using a simulated Nao platform, a simulated EMG signal, and a simulated human feedback model. We have previously shown the performance of this experimental set-up to be comparable between simulation and real-world experiments \cite{Mathewson2016SimultaneousLearning}. By simulating the human feedback, we are able to characterize and vary important latent variables hidden from the agent which impact the learning of the system. For this study, we investigate: the rate at which a human-delivered feedback should decay ($smear$), the probability with which the human will provide a feedback ($P(feedback)$), and the probability that this feedback will be correct for a given MDP ($P(correct)$). These are critical variables that have been estimated in previous experiments \cite{Knox2015FramingPerformance,Loftin2016LearningLearning}, we aim to improve understanding of their impact through an experimental grid sweep over the variables of interest and investigation into the results.

\section{Experiments}
We extend on the results in \cite{Mathewson2016SimultaneousLearning} by exploring the impacts of varying model parameters of human trainer feedback on the RL system during the performance of a self-mirrored movement control task. In this task, we preprogram the left arm of the Nao to move in a periodic pattern of flexion and extension at the elbow joint. The RL agent controls the right arm and selects angular displacement actions attempting to match the pattern of the left. With this configuration we are able to define an optimal policy, which would track the pre-programmed arm exactly, with this optimal trajectory we are able to derive MDP reward given a set angular error threshold. When the RL-controlled elbow joint is within the angular deviation threshold of the preprogrammed elbow joint then a reward of 1 is received from the MDP, otherwise, a negative relative error is delivered proportional to the difference between the actual and optimal angles. 

We are interested in modelling \textit{smear}, the time-decay with which the feedback given by the human should be decayed. As the human is unable to give feedback at every step that an agent takes, we need to account for the fact that after the exact time step a feedback is given there are likely suboptimal states which support the optimal trajectory. With a decay parameter we are able to \textit{smear} the human feedback forward in time, it has been shown that the limited human feedback can be applied across near-optimal state-action pairs, and support the agent learning an optimal solution \cite{Pilarski2011OnlineLearning}. We further explore the following characteristics of human feedback: ($P(feedback)$) the probability of giving feedback on each time step, and ($P(correct)$) the probability of giving correct vs. incorrect feedback. These are important latent human parameters to understand, cognitively they represent how effective and attentive a human trainer is. 

The continuous state space is defined by the filtered, time-averaged, and dimensionally reduced EMG signal and the angle of the actuated joint, and is represented with approximation using tile coding  \cite{Mathewson2016SimultaneousLearning}. Parameters were set as follows: $\alpha_v = 0.1/m$, $\alpha_\mu = \alpha_\sigma$, $\gamma = 0.9$, $\lambda_w = 0.3$, $\lambda_v = 0.7$, joint angles were limited by manufacturer specifications at $\theta \in [0.0349, 1.5446]$ rads. Weight vectors $\mathbf{w}_\mu, \mathbf{w}_\sigma, \mathbf{v}, \mathbf{e}_\mu, \mathbf{e}_\sigma, \mathbf{e}_\mathbf{v}$ were initialized to $\mathbf{0}$ and standard deviation was bounded by $\sigma \geq 0.01$. The eligibility trace update for the critic is scaled by $\gamma$ to speed up convergence. Maximum number of time steps = 10k, learning update and action selection occurred at ~33 Hz or every ~30 ms, and angular deviation threshold was set to $\Delta\theta_{max} = 0.1$, absolute angular joint updates were clipped to 0.1 and actions were selected and performed on every time step.

The ACRL system was trained online with simulated human feedback and simulated EMG control signals (designed to mimic acceptable control signals). Human feedback is integrated into the learning algorithm as reward accumulated on Step 6 of Algorithm \ref{alg:cacrl}. Performance was measured by taking the average mean absolute angular error from the last 5k steps. This was done to compare the experimental results after some learning and helped to reduce noise intrinsic in early learning.

This paper presents results of a parameterized grid sweep over three parameters with given estimates of reasonable values: $smear = (0.2, 0.5, 0.9)$, $P(feedback) = (0.03, 0.05, 0.09)$, $P(correct) = (0.6, 0.75, 0.9)$. Additionally, as a control case, $n = 60$ trials without human-feedback were performed. On all time steps MDP reward and human reward were directly summed and applied to the learning agent update (Algorithm \ref{alg:cacrl}).

\section{Results}
\begin{figure}[h!]
	\centering
    \includegraphics[width=3.3in]{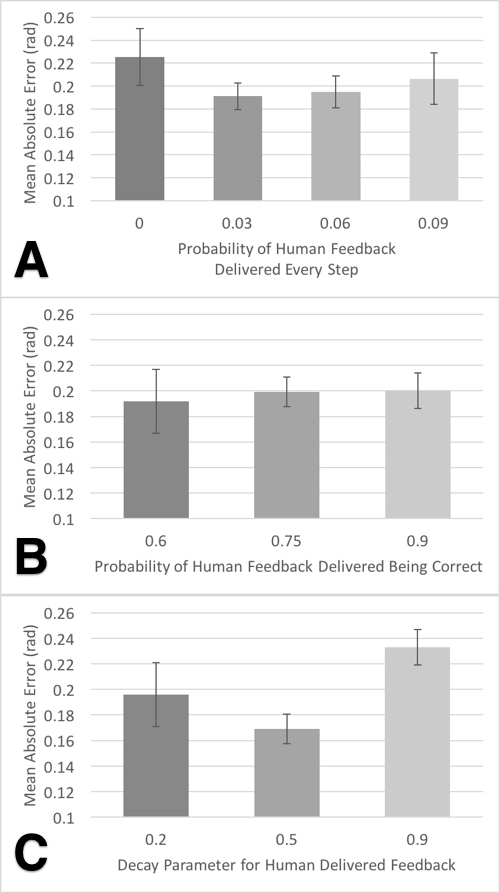}
  	\caption{Mean and standard error over experimental conditions A) $P(feedback)$, B) $P(correct)$, C) $smear$.}
    \label{fig:fig2}
\end{figure}

The results are presented in \ref{fig:fig2}. Results are presented which show performance over a variety of combinations of parameters for the latent variables of interest: $P(feedback)$, $P(correct)$ and $smear$. Results indicate that human interaction improves agent performance on a self-mirroring movement task where performance is measured by the mean angular error over the last 5k time steps. Fig. \ref{fig:fig2}A shows that a lower probability of potentially incorrect feedback provides better performance. Fig. \ref{fig:fig2}B shows that there may not be a significant difference in performance when varying the probability of the correctness human feedback, given tested values of $P(feedback)$. This may also be due to the tested values, which were all greater than a 50\% chance of being correct. Fig. \ref{fig:fig2}C shows that there is a benefit to selecting a \textit{smear} decay value appropriate for the task and robotic control system, this parameter may vary task to task and care must be taken when selecting the \textit{smear} constant. The results indicate that there is benefit to be gained by correct modelling the latent variables associated with human reward signal to allow for true simultaneous incorporation of human control and feedback. These results indicate that the ACRL algorithm robust to a small amount of incorrect feedback.

On average without human-feedback the RL agent was able to attain a mean absolute error on the final 5k steps of $0.22 \pm 0.02$ (mean $\pm$ SEM, n=60). In comparison, the optimal set of parameters $(P(feedback)=0.06, P(correct)=0.6, smear=0.5)$ was able to attain a performance of $0.12 \pm 0.01$ (mean $\pm$ SEM, n=7), the worst performing set of parameters $(P(feedback)=0.09, P(correct)=0.9, smear=0.9)$ attained a performance of $0.38 \pm 0.18$ (mean $\pm$ SEM, n=4). A total of 232 trials were run over parameter combinations.

\section{Discussion}
The experiments in this paper are performed using a simulated Nao, simulated EMG signal and simulated human feedback. It has been previously shown the performance of this experimental set-up to be comparable between simulation and real-world experiments \cite{Mathewson2016SimultaneousLearning}. In this related work we explore the degree to which the learning system is affected when incorporating real human feedback. While working in simulation allows rapid iteration and enables testing of many different algorithmic characteristics, simulation is often an easier learning problem than the real-world, due to simplified physics and reduced noise. Future work will address robust modelling real human feedback, and quantify impact of varying feedback density and correctness. We have shown that \textit{smearing} of human feedback impacts learning, future work will investigate if the decay of human delivered rewards is best modelled as time dependent over task performance and if optimal decay parameters may be learned online.

In this paper we found that modelling the delivery of human feedback can significantly impact the performance of an ACRL algorithm. While we have not optimized for the human feedback characteristics, these results indicate that some human reward paradigms may be preferable to others \cite{Loftin2016LearningLearning}. This idea is explored in \cite{Macglashan2016ConvergentHumans} where modelling the user feedback as an advantage function, we can understand positive feed back as \textit{'yes, that was good'} and negative feedback as \textit{'no, that was bad'}. A greater understanding of human reward strategies is required. Personalized robotics will demand perception of human strategies to learn optimal in a very few sample. Future work will focus predicting and optimizing for known and uncertain feedback strategies.

Linking control signals in state space with feedback shaping reward signals effectively blends multisensory human data to the learning agent. There remains an open problem of how feedback should best be interpreted by the learning agent and how to encourage human feedback without causing prohibitive additional cognitive load. Modelling, and predicting, human feedback may relieve burden while allowing for shaping control signal interpretation. Human feedback is beneficial to the agent, providing it adds complimentary information about the contextual state the agent is in. Human feedback may shape the MDP reward with more specificity and more often than the sparse, delayed, MDP-derived reward. 

Our results demonstrate potential benefits by introducing well modelled human feedback into the robotic learning system. The inclusion of human shaping signals was shown to improve performance over strictly environmentally derived reward. Providing consistent, correct feedback demands cognitive attention from the user which may be difficult if the user is also required to provide control signals to the robotic system. Future work is introduced to explore implications of inviting humans to simultaneously provide control and feedback signals to learning systems.

\section{Conclusions}
This paper contributes a set of results from experiments incorporating simulated human feedback and simultaneous human control in the training of a semi-autonomous robotic agent. These results indicate that task performance increases with the incorporation of human feedback into existing actor-critic reinforcement learning algorithms. These results support the idea that human interaction can improve performance in complex robotic tasks when the human feedback is delivered correctly, consistently, and on a time scale consistent with the original learning problem.

This work supports an emerging viewpoint surrounding human training of a robotic system tightly coupled to a user. By showing improving the performance of the RL agent this work further supports the sharing of autonomy between human and machine.

\section{Acknowledgments}
This work is supported by National Sciences and Engineering Research Council of Canada (NSERC), Alberta Innovates – Technology Futures (AITF), the Canada Research Chairs Program, the Canada Foundation for Innovation, and the Alberta Machine Intelligence Institute (Amii)/Alberta Innovates Centre for Machine Learning (AICML).

\bibliography{Mendeley_AAAI-SS-2017.bib}
\bibliographystyle{aaai}
\end{document}